\documentclass[a4paper,fleqn]{cas-dc}

\usepackage[numbers]{natbib}
\usepackage{mhchem}
\usepackage{booktabs}
\usepackage{threeparttable}

\def\tsc#1{\csdef{#1}{\textsc{\lowercase{#1}}\xspace}}
\tsc{WGM}
\tsc{QE}

\begin{document}
\let\WriteBookmarks\relax
\def\floatpagepagefraction{1}
\def\textpagefraction{.001}

\shorttitle{Implementation Pitfalls for Carbonate Mineral Dissolution -- a Technical Note}    

\shortauthors{F.J. Weiss, L. Keim, K. Wendel, H. Class}  

\title [mode = title]{Implementation Pitfalls for Carbonate Mineral Dissolution -- a Technical Note}  



%

\author[1]{Fiona J. Weiss}[orcid=0000-0002-0852-1500]
\ead{fiona.weiss@iws.uni-stuttgart.de}
\cormark[1]




\credit{Simulation, Visualization, Writing (original draft, review and editing)}

\affiliation[1]{organization={Institute for Modeling Hydraulic and Environmental Systems, University of Stuttgart},
            city={70569 Stuttgart},
            country={Germany}}

\author[1]{Leon Keim}[orcid=0000-0003-2456-1809]
\credit{credit Leon}
\author[1]{Kai Wendel}[orcid=0000-0003-1871-1540]
\credit{credit Kai}
\author[1]{Holger Class}[orcid=0000-0002-4476-8017]
\credit{credit Holger}






\cortext[1]{Corresponding author}



\begin{abstract}
In reactive transport modeling, an accurate understanding of reaction rates is essential; discrepancies in parameter reporting can greatly affect simulation results. This technical note identifies an issue with the reporting of rate parameters for carbonate mineral dissolution in a widely used database for reactive transport modeling based on \citet{Palandri2004}. Specifically, the reaction order was reported with respect to the partial pressure $P(\ce{CO2})$ rather than the activity of $\ce{H2CO3}^\ast$, causing a considerable overestimation of reaction timescales. We demonstrate the implications of this error by simulating a calcite dissolution batch experiment using Reaktoro and DuMuX and comparing the results to experimental data. By adjusting the parameter to align with established literature, we demonstrate an improved fit between simulated and experimental data. Discrepancies in reaction timescales were reduced by an order of magnitude, emphasizing the importance of regular validation of simulations with experimental data. \nocite{*}
\end{abstract}





\maketitle

\section{Introduction}\label{sec: Introduction}
Chemical equilibrium often cannot be assumed in systems with slow reaction kinetics such as mineral dissolution processes, thus understanding the rates of mineral dissolution and precipitation is essential. Particularly for processes like CO$_2$ sequestration in geologic formations, where injected CO$_2$ disturbs the system far from equilibrium, the slow kinetics of mineral-CO$_2$-water interactions dictate the re-equilibration timescales and the rates at which primary minerals dissolve and secondary minerals precipitate.

A commonly used reaction rate model, along with the necessary parameters for minerals frequently studied, can be found in a USGS report by \citet{Palandri2004}. This report serves as the foundation for a widely used database for computational programs in the reactive transport modeling community. The parameters for the \citet{Palandri2004} model were derived from experimental studies reported in older literature. However, we identified an issue with the reporting of rate parameters for carbonate minerals. Specifically, the authors reported the reaction order $n$ of the carbonate mechanism with respect to the partial pressure $P(\ce{CO2})$, when it should have been reported with respect to the activity of $\ce{H2CO3}^\ast$, to be consistent with the literature used to derive these parameters \cite{Busenberg1982, Chou1989, Johnson2001, Plummer1978, Pokrovsky1999, Talman1990}. This oversight leads to a considerable overestimation of reaction timescales. To address this issue, we propose an adjustment of the relevant section of the \citet{Palandri2004} report.

\section{Methods}\label{sec: Methods}
For carbonate minerals, three elementary mechanisms are involved in the dissolution reaction \cite{Chou1989, Plummer1978}; namely the acid mechanism, the carbonate mechanism, and the neutral mechanism with forward reaction rate parameters $k_1$, $k_2$, and $k_3$ and backward reaction rate parameters $k_{-1}$, $k_{-2}$, and $k_{-3}$, respectively.
\begin{equation} 
    \begin{split}
        \ce{&MCO3 + H^+ <=>[k_1][k_{-1}] M^{2+} + HCO3^-} \\
        \ce{&MCO3 + H2CO3}^\ast \ce{ <=>[k_2][k_{-2}] M^{2+} + 2HCO3^-} \\
        \ce{&MCO3 <=>[k_3][k_{-3}] M^{2+} + CO3^{2-}} 
    \end{split}
\end{equation}
where $\ce{M}$ represents the metal ion. The total forward and backward reaction rates $R_f$ and $R_b$ are obtained as
\begin{equation}\label{eq2}
    \begin{split}
        R_f &= k_1 \, a_{\ce{H^+}} + k_2 \, a_{\ce{H2CO3}^\ast} + k_3 \\
        R_b &= k_{-1} \, a_{\ce{M^{2+}}} \, a_{\ce{HCO3^-}} + k_{-2} \, a_{\ce{M^{2+}}} \, a_{\ce{HCO3^-}}^2 + k_{-3} \, a_{\ce{M^{2+}}} \, a_{\ce{CO3^{2-}}} \\ 
        &= k_4 \, a_{\ce{M^{2+}}} \, a_{\ce{HCO3^-}}
    \end{split}
\end{equation}

The empirical rate parameters $k_1$, $k_2$, $k_3$, and $k_4$ are either stated in literature directly \cite{Chou1989}, or are described by the equation
\begin{equation}\label{eq3}
    \log \, k_i = a_i\, (1/T) + b_i
\end{equation}
where $T$ is the temperature, and $a_i$ and $b_i$ can be found in literature \cite{Busenberg1982, Plummer1978}.

A semi-empirical rate model that can fit the experimental data reasonably well without explicit knowledge of the backward rate coefficients is given in \citet{Palandri2004} as
\begin{equation} \label{eq4}
    \begin{split}
        R = -A & \left[k_{\text{acid}}^{298.15\text{K}}\, e^{-\frac{E_1}{R} \left(\frac{1}{T} - \frac{1}{298.15\text{K}}\right)} \, a_{\ce{H^+}}^{n_1} \, \left(1-\Omega^{p_1}\right)^{q_1} \right. \\
        & \left.+ k_{\text{neutral}}^{298.15\text{K}}\, e^{-\frac{E_2}{R} \left(\frac{1}{T} - \frac{1}{298.15\text{K}}\right)} \, \left(1-\Omega^{p_2}\right)^{q_2} \right. \\
        & \left.+ k_{\text{carbonate}}^{298.15\text{K}}\, e^{-\frac{E_3}{R} \left(\frac{1}{T} - \frac{1}{298.15\text{K}}\right)} \, a_{\ce{H2CO3}^\ast}^{n_3} \, \left(1-\Omega^{p_3}\right)^{q_3} \right] 
    \end{split}
\end{equation}
where $A$ is the mineral surface area, $E_i$ is the activation energy of mechanism $i$, $R$ is the gas constant, $T$ is the temperature, $a_j$ is the activity of species $j$, and $\Omega$ is the mineral saturation index. $n_i$, $p_i$, and $q_i$ are dimensionless empirical parameters. Note that for 25°C, the exponential term becomes unity and the relation for the forward reaction rate \eqref{eq2} can be obtained. Also note that the indices of the neutral mechanism and of the carbonate mechanism are switched in \eqref{eq4}.

The saturation index $\Omega$ is computed using the activity of the mineral species: DuMuX assumes an ideal activity model, whereas Reaktoro uses the Pitzer model for activities.

The rate parameters $k_{\text{acid}}^{298.15\text{K}}$, $k_{\text{neutral}}^{298.15\text{K}}$, and $k_{\text{carbonate}}^{298.15\text{K}}$ can be obtained from literature \cite{Busenberg1982, Chou1989, Johnson2001, Plummer1978, Pokrovsky1999, Talman1990}; however, a unit conversion may be necessary. As an example, we demonstrate this derivation for calcite. \cite{Plummer1978} yields the reaction orders $n_1=1$ with respect to $\ce{H+}$ for the acid mechanism and $n_3=1$ with respect to $\ce{H2CO3}^\ast$ for the carbonate mechanism, as well as the empirical rate parameters $a_1 = -444.0$, $b_1=0.198$, $a_2 = -2177.0$, $b_2 = 2.84$, $a_3 = -317.0$, and $b_3= -5.86$; thus, using equation \eqref{eq3} and $T = 298.15$K
\begin{equation}
    \begin{split}
        &\log\, k_1 = -1.3 \Rightarrow k_1 = 10^{-1.3} \,\frac{\text{mmol}}{\text{cm}^2\,\text{s}} = 10 ^{-0.3} \,\frac{\text{mol}}{\text{m}^2\,\text{s}}\\
        &\log\, k_2 = -4.46 \Rightarrow k_2 = 10^{-4.46} \,\frac{\text{mmol}}{\text{cm}^2\,\text{s}} = 10 ^{-3.46} \,\frac{\text{mol}}{\text{m}^2\,\text{s}} \\
        &\log\, k_3 = -6.92 \Rightarrow k_3 = 10^{-6.92} \,\frac{\text{mmol}}{\text{cm}^2\,\text{s}} = 10 ^{-5.92} \,\frac{\text{mol}}{\text{m}^2\,\text{s}}
    \end{split}
\end{equation}
These values for $k_1$, $k_2$, and $k_3$ align reasonably well with $\log \, k_{\text{acid}}^{298.15\text{K}} = -0.30$, $\log \,k_{\text{carbonate}}^{298.15\text{K}} = -3.48$, and $\log \, k_{\text{neutral}}^{298.15\text{K}} =-5.81$ stated in \citet{Palandri2004}. The method can be applied similarly to the other carbonate minerals mentioned herein.

\section{Demonstration and Validation}\label{sec: Demonstration and Validation}
To illustrate the impact of using the model with parameters as stated by \citet{Palandri2004}, or any database implementing this data, we simulated a simple calcite dissolution batch test based on the experiment conducted in \citet{Plummer1976}. The test was simulated using Reaktoro \cite{leal2015reaktoro}, a commonly used computational tool for modeling chemically reactive systems, and the results were compared to the original experimental data from run 7 in \citet{Plummer1976}. 

In this experiment, conditions were set to 25°C and 1 atm pressure. The solution consisted of 500 mL distilled water through which CO$_2$ was bubbled at a rate of 2.0 L/min. A total of 11.748 g of calcite was added, corresponding to a total surface area of 5568 cm$^2$. The surface area was estimated from the weight of the sized material, assuming each particle to be a rhombohedron of width equal to the mean sieve size, as described in the original paper. Because the weight of calcite exceeded its solubility in CO$_2$-saturated water by at least five times (and in most cases more than twenty times), the experiments could assume a constant surface area throughout.

\begin{figure} 
  \centering
    \includegraphics[scale=1.0]{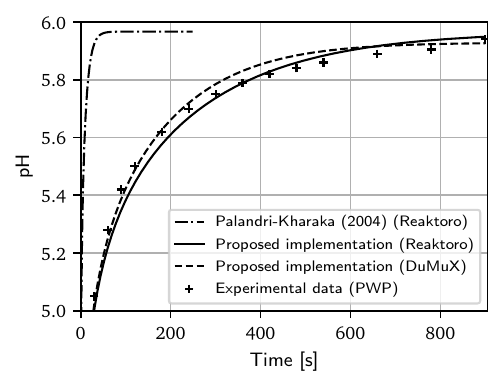}
    \caption{Comparison of the originally reported values in \citet{Palandri2004}, our proposed adjustment, simulation of the model from \citet{Plummer1978} in DuMuX and experimental data from \citet{Plummer1976}. Evidently, reaction time scales are significantly overestimated by the original report.}
    \label{fig1}
\end{figure}

Figure \ref{fig1} shows that implementing the reaction order $n$ of the carbonate mechanism with respect to $P(\ce{CO2})$ results in a considerable overestimation of reaction timescales; the time to reach saturation is approximately 20 times faster than observed experimentally. Conversely, implementing the reaction order $n$ with respect to $\ce{H2CO3}^\ast$ aligns well with the experimental data, demonstrating the validity of the proposed adjustment. Similar observations can be made for the species molality of $\ce{Ca^{2+}}$ in water.

Implementing the original model from \cite{Plummer1978} in the software DuMuX \cite{Koch2021} provides additional support for our claim. This approach also yields results consistent with our proposed implementation in Reaktoro and experimental data. The slight difference in outcomes between Reaktoro and DuMuX can be explained by the different underlying approaches; Reaktoro's kinetics solver minimizes the Gibbs energy, while DuMuX solves the electroneutrality equation.

\section{Conclusion}\label{sec: Conclusion}
This technical note demonstrates that the original parameters reported in \citet{Palandri2004}, using the reaction order $n$ with respect to $P(\ce{CO2})$, very significantly overestimate reaction timescales. Adjusting the reaction order to reference $\ce{H2CO3}^\ast$ is required and aligns simulation results with experimental data, as confirmed through simulations in Reaktoro and DuMuX.

The corrected kinetic parameters summarized in Table \ref{tab:kinetic_params} provide a reference for future simulations using carbonate minerals.

\begin{table*}[ht]
	\begin{threeparttable}
		\caption{Kinetic parameters for carbonate mineral dissolution: rate constant $\log k$ [mol/(m$^{2}\,$s)], activation energy $E_a$ [kJ/mol], and reaction order $n$ [-] for the acid, neutral, and carbonate mechanisms, with corresponding sources.}
		\label{tab:kinetic_params}
		\begin{tabular*}{\textwidth}{@{\extracolsep{\fill}}lccccccccl}
			\toprule
			Mineral & \multicolumn{3}{c}{Acid mechanism} & \multicolumn{2}{c}{Neutral mechanism} & \multicolumn{3}{c}{Carbonate mechanism} & Source \\
			\cmidrule(lr){2-4} \cmidrule(lr){5-6} \cmidrule(lr){7-9}
			& $\log k$ & $E_a$ & \tnote{a} $n$ & $\log k$ & $E_a$ & $\log k$ & $E_a$ & \tnote{b} $n$ &  \\
			\midrule
		Calcite    						& -0.30 & 14.4 & 1.000 & -5.81 & 23.5 & -3.48 & 35.4 & 1.000 & \cite{Plummer1978, Talman1990} \\
		Dawsonite  						&   --  &  --  &  --   & -7.00 & 62.8 &   --  &  --  &  --   & \cite{Johnson2001} \\
		Sedimentary (disordered) Dolomite  & -3.19 & 36.1 & 0.500 & -7.53 & 52.2 & -5.11 & 34.8 & 0.500 & \cite{Busenberg1982} \\
		Hydrothermal (ordered) Dolomite    & -3.76 & 56.7 & 0.500 & -8.60 & 95.3 & -5.37 & 45.7 & 0.500 & \cite{Busenberg1982} \\
		Magnesite 						& -6.38 & 14.4 & 1.000 & -9.34 & 23.5 & -5.22 & 62.8 & 1.000 & \cite{Chou1989, Pokrovsky1999} \\
		\bottomrule
	\end{tabular*}
\begin{tablenotes}
	\item[a] Reaction order $n$ with respect to $\ce{H+}$.
	\item[b] Reaction order $n$ with respect to $\ce{H2CO3}^\ast$.
\end{tablenotes}
\end{threeparttable}
\end{table*}

We encourage geochemical modelers to pay special attention to how reaction rates are defined in embedded kinetics solvers and databases used by their simulation software. Errors in parameter reporting can lead to false simulation results; therefore, regular validation of simulations with experimental data is important to ensure reliable results.

\section{Data Availability Statement}\label{sec: Data Availability Statement}
The source code used for the numerical simulations is available in \cite{weis2024implementation}. The postprocessing script is also provided in \cite{weis2024implementation}. A precompiled Docker image for running the simulations is accessible at \cite{darus-code}, and all files required for postprocessing can be downloaded from \cite{darus-data}.

\section{Funding}\label{sec: Funding}
Open Access funding enabled and organized by Projekt DEAL. Funded by Deutsche Forschungsgemeinschaft (DFG, German Research Foundation) under Germany’s Excellence Strategy - EXC 2075 – 390740016. The work is further funded by Deutsche Forschungsgemeinschaft (DFG, German Research Foundation) with the following projects: "GAS-REACT: GAS interchange and REACTive processes in coupled subsurface/atmosphere systems(HE 9610/CL 190))" - project number 529041613 and "Experimental and numerical investigations on density-driven dissolution of CO2 and related carbonate dissolution in karst water" - project number 508470891. Funded by the Deutsche Forschungsgemeinschaft (DFG, German Research Foundation) - Project Number 327154368 (SFB 1313).









\bibliographystyle{cas-model2-names}

\bibliography{cas-refs}

@ARTICLE{Chou1989,
  author  = {Chou, L. and Garrels, R. M. and Wollast, R.},
  title   = {Comparative study of the kinetics and mechanisms of dissolution of carbonate minerals},
  journal = {Chemical Geology}, 
  volume  = {78},
  year    = {1989},
  pages   = {269-282}
}

@ARTICLE{Palandri2004,
  author  = {Palandri, J. L. and Kharaka, Y. K.},
  title   = {A compilation of rate parameters of water-mineral interaction kinetics for application to geochemical modeling},
  journal = {US Geological Survey Open File Report 2004-1068}, 
  year    = {2004}
}

@ARTICLE{Plummer1976,
  author  = {Plummer, L. N. and Wigley, T. M. L.},
  title   = {The dissolution of calcite in $\ce{CO_2}$-saturated solutions at 25{°C} and 1 atmosphere total pressure},
  journal = {Geochimica et Cosmochimica Acta}, 
  volume  = {49},
  year    = {1976},
  pages   = {191-202}
}

@ARTICLE{Plummer1978,
  author  = {Plummer, L. N. and Wigley, T. M. L. and Parkhurst, D. L.},
  title   = {The kinetics of calcite dissolution in $\ce{CO_2}$-water systems at 5° to 60°{C} and 0.0 to 1.0 atm $\ce{CO_2}$},
  journal = {American Journal of Science}, 
  volume  = {278},
  year    = {1978},
  pages   = {179-216}
}

@ARTICLE{Busenberg1982,
  author  = {Busenberg, E. and Plummer, L. N.},
  title   = {The kinetics of dissolution of Dolomite in $\ce{CO_2}$-$\ce{H_{2}O}$ systems at 1.5 to 65{°C} and 0 to 1 atm ${P_{\ce{CO_2}}}$},
  journal = {American Journal of Science}, 
  volume  = {282},
  year    = {1982},
  pages   = {45-78}
}

@ARTICLE{Pokrovsky1999,
  author  = {Pokrovsky, O. S. and Schott, J.},
  title   = {Processes at the magnesium-bearing carbonates/solution interface. {II}. {K}inetics and mechanism of magnesite dissolution},
  journal = {Geochimica et Cosmochimica Acta}, 
  volume  = {63},
  year    = {1999},
  pages   = {881-897}
}

@misc{leal2015reaktoro,
  author = {Leal, A. M. M.},
  title = {Reaktoro: An open-source unified framework for modeling chemically reactive systems},
  year = {2015},
  url = {https://reaktoro.org}
}

@article{Koch2021,
title={{DuMu$^\text{x}$ 3 -- an open-source simulator for solving flow and transport problems in porous media with a focus on model coupling}},
journal  = {Computers \& Mathematics with Applications},
volume = {81},
pages = {423-443},
year = {2021},
issn = {0898-1221},
doi = {https://doi.org/10.1016/j.camwa.2020.02.012},
author = {Timo Koch and Dennis Gläser and Kilian Weishaupt and Sina Ackermann and Martin Beck and Beatrix Becker and Samuel Burbulla and Holger Class and Edward Coltman and Simon Emmert and Thomas Fetzer and Christoph Grüninger and Katharina Heck and Johannes Hommel and Theresa Kurz and Melanie Lipp and Farid Mohammadi and Samuel Scherrer and Martin Schneider and Gabriele Seitz and Leopold Stadler and Martin Utz and Felix Weinhardt and Bernd Flemisch}
}

@inproceedings{Johnson2001,
  author    = {Johnson, J. W. and Nitao, J. J. and Steefel, C. I. and Knauss, K. G.},
  title     = {Reactive transport modeling of geologic $\ce{CO_2}$ sequestration in saline aquifers:
               the influence of intra-aquifer shales and the relative effectiveness of structural,
               solubility, and mineral trapping during prograde and retrograde sequestration},
  booktitle = {First National Conference on Carbon Sequestration},
  year      = {2001},
  address   = {Washington, D.C.},
  month     = may,
  pages     = {1--60},
  note      = {May 14--17, 2001}
}

@incollection{Talman1990,
  author    = {Talman, S. J. and Wiwchar, B. and Gunter, W. D.},
  title     = {Dissolution kinetics of calcite in the $\ce{H_{2}O}$-$\ce{CO_2}$ system along the steam saturation curve to 210°{C}},
  booktitle = {Fluid-Mineral Interactions: A Tribute to H. P. Eugster},
  editor    = {Spenser, R. J. and Chou, I.-M.},
  publisher = {Lancaster Press, Inc.},
  address   = {San Antonio, Texas},
  year      = {1990},
  pages     = {41--55}
}

@data{darus-data,
author = {Keim, Leon and Weiss, Fiona J. and Wendel, Kai and Class, Holger},
publisher = {DaRUS},
title = {{Replication Data for: Implementation Pitfalls for Carbonate Mineral Dissolution – a Technical Note}},
UNF = {UNF:6:HGhJePghgVQkheBN8H3abw==},
year = {2025},
version = {DRAFT VERSION},
doi = {10.18419/darus-4715},
url = {https://doi.org/10.18419/darus-4715}
}

@data{darus-code,
author = {Keim, Leon and Weiss, Fiona J. and Wendel, Kai and Class, Holger},
publisher = {DaRUS},
title = {{Replication Code for: Implementation Pitfalls for Carbonate Mineral Dissolution – a Technical Note}},
year = {2025},
version = {DRAFT VERSION},
doi = {10.18419/darus-4716},
url = {https://doi.org/10.18419/darus-4716}
}

@software{weis2024implementation,
author = {Weiss, Fiona J. and Keim, Leon and Wendel, Kai and Class, Holger},
title = {Implementation Pitfalls for Carbonate Mineral Dissolution},
year = {2024},
institution = {Institute for Modelling Hydraulic and Environmental Systems},
url = {https://git.iws.uni-stuttgart.de/dumux-pub/weis2024a},
version = {1.0.0} 
}



\end{document}